\documentstyle[12pt,amssymbols]{article}
\input math_macros.tex

\begin{document}
\begin{titlepage}
\begin{center}
October 1994          \hfill LBL-37560 \\
          \hfill    UCB-PTH-95/28 \\
         \hfill    hep-th/9507156 \\

\vskip .25in

{\large \bf Convergent sequences of perturbative approximations for the
anharmonic oscillator\\
II. Compact time approach}\footnote{This work was supported in part
by the Director, Office of
Energy Research, Office of High Energy and Nuclear Physics, Division of
High Energy Physics of the U.S. Department of Energy under Contract
DE-AC03-76SF00098 and in part by the National Science Foundation under
grant PHY-90-21139.}

\vskip .25in

B. Bellet \\
P. Garcia\\

{\em Laboratoire de Physique Math\'ematique\\
Universit\'e de Montpellier II-CNRS\\
34095 Montpellier c\'edex 05}\\

and\\

A. Neveu\footnote{neveu@lpmsun2.lpm.univ-montp2.fr. On sabbatical leave
after Sept. 1$^{\hbox{st}}$ 1994 from Laboratoire de Physique Math\'ematique,
Universit\'e de Montpellier II-CNRS, 34095 Montpellier c\'edex 05}

{\em Theoretical Physics Group\\
    Lawrence Berkeley Laboratory\\
      University of California\\
    Berkeley, California 94720}

\end{center}

\vskip .5in

\begin{abstract}
We present an alternative pathway in the application of the variation
improvement of ordinary perturbation theory exposed in [1] which can preserve
the internal symmetries of a model by means of a time compactification.

\end{abstract}
\end{titlepage}
\renewcommand{\thepage}{\roman{page}}
\setcounter{page}{2}
\mbox{ }

\vskip 1in

\begin{center}
{\bf Disclaimer}
\end{center}

\vskip .2in

\begin{scriptsize}
\begin{quotation}
This document was prepared as an account of work sponsored by the United
States Government. While this document is believed to contain correct
 information, neither the United States Government nor any agency
thereof, nor The Regents of the University of California, nor any of their
employees, makes any warranty, express or implied, or assumes any legal
liability or responsibility for the accuracy, completeness, or usefulness
of any information, apparatus, product, or process disclosed, or represents
that its use would not infringe privately owned rights.  Reference herein
to any specific commercial products process, or service by its trade name,
trademark, manufacturer, or otherwise, does not necessarily constitute or
imply its endorsement, recommendation, or favoring by the United States
Government or any agency thereof, or The Regents of the University of
California.  The views and opinions of authors expressed herein do not
necessarily state or reflect those of the United States Government or any
agency thereof, or The Regents of the University of California.
\end{quotation}
\end{scriptsize}

\vskip 2in

\begin{center}
\begin{small}
{\it Lawrence Berkeley Laboratory is an equal opportunity employer.}
\end{small}
\end{center}

\newpage
\renewcommand{\thepage}{\arabic{page}}
\setcounter{page}{1}


\input epsf.tex
\parskip 10pt
\setcounter{equation}{0}

\section{Introduction.}
In the preceding paper \cite{bgn}, we presented a study
of large orders of a variational method which
provides convergent sequences of approximations
for the quantum anharmonic oscillator.
In this method, we introduced a mass term (with a
mass $\omega$) in the free Lagrangian and
subtracted it in the interaction part.
Then, we optimized at any order in terms
 of this variational parameter $\omega$
to get approximate values of different
 quantities.

We have in perspective the
use of such a method in quantum field theory.
For fermions like those of the Gross-Neveu model \cite{agn},
or QCD quarks, introduction of a mass
term causes {\em a priori} no problem, but for gluons
it would break gauge invariance, and there is no obvious
modification of the momentum dependent part of the
gluon propagator and the structure of the ultraviolet
infinities of the theory. However, a space
 compactification might provide some
good approximation of quark confinement, like in
the M.I.T. bag model \cite{MIT}.

In the M.I.T. bag model, the bag is a fundamental ingredient
of the theory, which hence is not QCD, while we would
advocate here an approach in which it would be added
to the free gluon and quark action in the form of
a modification of their propagators, the subtraction of
this modification being done in perturbation theory.
While formally when going to all orders one would have
done nothing, and still be dealing with QCD, a finite
given order would have to be optimized in the size of
the bag as the closest one could get to the infinite order
case. The advantage would be that different physics
would be involved, hopefully closer to that of the
full theory, with confinement, etc\ldots

Compactification as an intermediate step in calculations
in infrared divergent field theories has already been
considered by several authors \cite{LF} , with
reasonable success. These authors usually treat the
theory in a spatial box of length $l$, and extrapolate
the results to $l \rightarrow \infty$. We are here
advocating a different, if related, strategy, in which
the box is in some sense subtracted perturbatively, and
the calculation at a given order in perturbation theory
optimized with respect to the size of the box, this being
supposed to mimick as well as it can the true all-orders
theory, where formally there is no dependence at all
on the size of the box.

In this paper, we try out this idea on the case of
the anharmonic oscillator, choosing an action with
compactified time as our starting point.
 We find a configuration for
the optima and the approximate values of the
 ground state energy very similar to those
 of the harmonic approach of this problem
\cite{bgn}. In the second section, we study
the interaction part of the action to be used
in order to formally leave the total action of a pure
 anharmonic oscillator unchanged.
We make a choice which allows relatively simple
 perturbative calculations starting from a
theory governed by a compactified action.
The perturbation theory for such
a compactified action is considered
in the third section. In the
fourth section, we establish the
variational-perturbative expansion of
the mean value of the Hamiltonian and
optimize it in terms of our variational
 parameter (the size of the box) up to
order $16$. We interpret the configuration
 of the extrema using the large order
behavior of the expansion, which is qualitatively
much the same as in the harmonic approach of the
previous paper, to which we  compare our results.
They turn out to be less accurate numerically for
similar orders in the approximation, which
reflects the fact that approximating an anharmonic
potential by a compact time dimension is much worse
than approximating it by a harmonic potential. Nevertheless,
the convergence properties are qualitatively very similar.

 \section{Time compactification
of a pure anharmonic oscillator.}

The Euclidean action for a pure
 anharmonic oscillator writes:
\begin{equation}S\;=\;\int^{\infty}
_{-\infty} \Bigl[\;{1\over 2} \left({\partial_{t}\phi}
\right)^2+{\lambda\over 4} \phi^4\; \Bigr]\; dt\;.
\end{equation}

Severe infrared divergences prevent any perturbation
in powers of $\lambda$. A way to tame these divergences
is of course to introduce an $\omega^2\phi^2$ term
as was done in the previous paper. Here, we would
instead like to replace in some sense the free term of
the action of equation (1) by the compactified
free action $S^0(\tau)$:

\begin{equation}S^0(\tau)\;=\;
\int^{{\pi\over 2}\tau}_{-{\pi\over 2}\tau}
{1\over 2}\left({\partial_{t}\phi}\right) ^2\;
 dt\;,
\end{equation}
and rewrite the total action $S$ as:
\begin{equation}
S=S^0(\tau)+S^I(\tau).
\end{equation}
Then, we will be able to perform a perturbative expansion
 of any quantity in powers of the interaction
 term $S^I(\tau)$ around the free term
 $S^0(\tau)$. To do so, let us consider
 the Euclidean action $S_T$ of a time
compactified pure anharmonic oscillator

\begin{equation}
 S_T=\int^{{\pi\over2}T}_{-{\pi\over2}T}
[\;{1\over2} \left({\partial_{t}\phi}\right)^2\;
+\;{\lambda\over4} \phi^4 \;]\;dt.
\end{equation}

With $S_T$, one can compute in perturbation theory
in powers of $\lambda$ at any given $T$ (provided
one uses appropriate boundary conditions, see below).
To see how $S$ and $S_T$ may be related, we perform
 the following change of variable:
\begin{equation}
x=T \tan{t\over T}\;\;;\;\;
 dx=(1+{x^2\over T^2})\; dt ,
\end{equation}
under which the action $S_T$ becomes:
\begin{equation}
 S_T=\int^{\infty}_{-\infty}[\;{1\over 2}
(1+{x^2\over T^2}) \left({\partial_{x}\phi}
\right)^2 \;+\;{\lambda\over4}{1\over(1+{x^2
\over T^2})} \phi^4 \;]\;dx.
\end{equation}
On the other hand, we can rewrite the action $S$ as
\begin{equation}
S\;=\;\int^{\infty}_{-\infty}\Bigl[\;{1\over 2} \;
(1+{x^2\over \tau^2}\;(1-\varepsilon))
\left({\partial_{x}\phi}\right)^2+{\lambda\over4}
{1\over{1+{x^2\over \tau^2}\;(1-\varepsilon)}}
\;\phi^4 \;\Bigr]\;dx,
\end{equation}
where, strictly speaking, $\varepsilon=1$. Comparing (7) with
 the expression of the compactified action (6),
we see how to relate $S$ and $S_T$: following the
procedure of the previous paper, one computes
the  quantity of interest to some order
in $\lambda$ using $S_T$, sets $T=\tau(1-\varepsilon)^{-1/2}$,
expands in powers of $\varepsilon$ to the same total order as in
$\lambda$, sets $\varepsilon=1$ and optimizes the
result with respect to $\tau$. This should provide
us with an approximate value of the same quantity
computed with $S$.

In more detail,  we can consider the
action (7) as:$$S\;=\;S^0(\tau)\;+\;S^I(\tau)
$$with $S^0(\tau)$ given by
\begin{equation}
S^0(\tau)\;=\;\int^{\infty}_{-\infty}{1\over2}\;
\Bigl( 1+{x^2\over \tau^2}\Bigr)\;\left({\partial_{x}\phi}
\right)^2\;dx
\end{equation}
and
\begin{equation}
S^I(\tau)\;=\;\int^{\infty}_{-\infty}\Bigl[\;-{1\over2}
\;\varepsilon\;{x^2\over \tau^2}\,
\left({\partial_{x}\phi}\right)^2\;+
\;{\lambda\over4}{1\over{1+{x^2\over \tau^2}\;
(1-\varepsilon)}} \;\phi^4\;\Bigr]\;dx.
\end{equation}

 The $p^{th}$ order in perturbation
 will contain all the terms proportional to
$\lambda^n\varepsilon^m$ with $n+m\leq p$.
 Note that in this approach, we do
not take into account the complete quartic term
of (4) but that we reconstruct it more and
more accurately as the
order of perturbation increases. This
feature does not seem to us to carry any pathology,
but simply follows from the choice of $S_T$
as our compactified action and the physics it contains.

This new choice for a free kinetic
term parameterized by $\tau$ in $S$ will
give rise for every finite order $p$ of perturbation
to an explicit dependence on $\tau$. Because the total
action $S$ remains unchanged at $\varepsilon=1$, one expects
this parameter $\tau$ to become irrelevant at infinite
order of perturbation. As already explained, this suggests to
fix this parameter by looking for the regions
 where the result to order $p$ is stationary in $\tau$.

Let us now discuss briefly the boundary conditions
in the action of equation (4).
  Periodic conditions $\phi_{-T}
=\phi_{T}$ are unsuitable for perturbation theory,
as the Gaussian kernel has an isolated zero
eigenvalue. Similarly for Neumann boundary conditions.
Hence, we shall use Dirichlet boundary conditions
in which

\begin{equation}
\phi_{\pm T}=0.
\end{equation}
which preserves the discrete $\phi \leftrightarrow -\phi$
symmetry. With this choice, all infrared problems
disappear at finite $T$.

In order to compare the results obtained by
this compact time method to those of the harmonic
 one \cite{bgn}, we would like to consider the
ground state energy. However, the method we
 described above implies calculation using
 a compactified action which is therefore not
invariant under translation in time. Then,
 the ground state energy is not a meaningful concept and
 we choose instead to consider the mean value of the
 Hamiltonian operator at some time $t_0$:
\begin{equation}
<\hat H_{t_0}>
\end{equation}
The variation over the size of the box $T$
provides some new approach of the limit
$T\rightarrow \infty$ (substituting
 $\tau\rightarrow T(1-\varepsilon)^{-1/2}$,
expanding in powers of $\varepsilon$ and
taking $\varepsilon=1$). In the limit of infinite
order of perturbation, the box should disappear
and we should recover the value of the ground state
energy when calculating $<\hat H_{t_0}>$.

\section{Perturbative expansion for
 the compactified action.}
In the Euclidean path integral representation, the
mean value of the Hamiltonian operator at time $t_0$
in the box writes:
\begin{equation}
 <\hat H_{t_0}>={<0;T\mid \hat H_{t_0}
 \mid 0;-T>\over<0;T\mid 0;-T>};
\end{equation}
where
\begin{eqnarray}
<0;T\mid \hat H_{t_0} \mid 0;-T>&=&
\int_{\phi_{\pm T}=0} D \phi\;\; H_{t_0}
 \;e^{- S_{T}[\phi]}\;,\nonumber\\<0;T
\mid 0;-T>&=&\int_{\phi_{\pm T}=0} D \phi
 \; e^{- S_{T}[\phi]}\;, \nonumber\\
- S_{T}[\phi]&=&- \int_{-T}^{T} \Bigl[ {1\over 2}
{(\partial_t\phi)}^2_t+{\lambda\over 4}
\phi^4_t\Bigr] \,dt \;,\nonumber\\ \mbox{and}
\hspace{3cm}H_{t_0}&=&-{1\over 2}
{(\partial_t\phi)}^2_{t_0}+{\lambda\over 4}
 \phi^4_{t_0}\;. \nonumber
\end{eqnarray}

\subsection{The partition function and
the mean value of the Hamiltonian.}
       Dividing by the partition
 function $Z(T)$ in (12) cancels all the
disconnected Feynman diagrams. We rewrite
 it:$$Z(T)=<0;T\mid 0;-T>=<0\mid \hat U_{2 T}
\mid 0> $$Here, in the Schr\"odinger
representation, the evolution operator
$\hat U_t$ of a quantum system described
 by $\hat H$ appears. It satisfies the
following equations:
\begin{equation}
 -{d\over dt}\hat U_t\;=\;\hat H\,
\hat U_t\;\;\; \mbox{and}\;\;\;\hat U_0
\;=\; 1.
\end{equation}
Then, taking the derivative of the partition function
with respect to the size of the box,
we get:

\begin{eqnarray}
{d\over dT}Z(T)&=&<0\mid{d\over dT}
 \hat U_{2 T} \mid 0> \nonumber\\
               &=&-2\, <0\mid
\hat H \hat U_{2 T} \mid 0> \nonumber\\
               &=&-2\, <0;T \mid \hat H_t
\mid 0;-T>, \nonumber
\end{eqnarray}
so that
\begin{equation}
{d\over dT}Z(T)=-2\;
 <-{1\over 2}\hat {(\partial_{t}\phi)}_t^2\;
+\;{\lambda\over 4}\hat \phi^4_t>;\;\;\forall\;
t\in [-T;T].
\end{equation}
Thus, the mean value of the
Hamiltonian operator does not depend of
the time when it is taken in the box. Moreover,
its perturbative expansion in powers of $T$ is
obtained from the one of $Z(T)$. However, the
variational procedure implies that at the order
 $p$ in perturbation, the $n^{th}$ order
contribution which is proportional to
 $\lambda^n\tau^{\nu}(1-\varepsilon)^{-\nu/2}$
 has to be expanded in powers of $\varepsilon$
up to order $p-n$. This $n^{th}$ power of
$\lambda$ does not take into account the
$\lambda$ present in front
of the quartic part of the Hamiltonian.
 But the two sides of equation (14)
identify themselves in perturbation, mixing
 that $\lambda$ with the perturbative one
\footnote{For example, in the perturbative
 expansion of $dZ(T)/dT$, the zeroth
order contribution of $<\hat \phi^4_t>$
is mixed with the first order of $<\hat
{(\partial_{t}\phi)}_t^2>$ .}.
So, we have to know kinetic and
potential contributions for each
order in the expansion of ${dZ(T)/dT}$.
 Nevertheless, we will use this equality
to compute at any order the value of
$<\hat {(\partial_{t}\phi)}_t^2>$ in
 term of $<\hat \phi^4_t>$ and $Z(T)$.
We decided to compute the mean value
of the Hamiltonian where its kinetic
and potential contributions do not
much vary and are the least dependent
on the size of the box, $i.e.$ in the
middle of the box at $t=0$.

\subsection{The perturbative
calculation at any order.}

 In order to develop a feeling on how the variational
 method works, we have to reach high
orders of perturbation, $i.e.$ calculate
a large number of coefficients. The Feynman diagrams
approach implies the calculation of
an increasing number of graphs. Instead, we
prefer a perturbative expansion of
 the evolution operator closer to the usual
Rayleigh-Ritz method. Expanding $\hat U_t$
in powers of the coupling constant
 $\lambda$, we get the following
expansions for $Z(T)$, $<\hat{\phi}^4_0>$
and $<\hat{(\partial_t \phi)}^2_0>$:
\begin{eqnarray}
<0;T|0;-T>&=& {1\over \sqrt{T}}\;\sum_{n=0}^{\infty}\;
({-\lambda/ 4})^n T^{3 n}\;\; {\bf Z}_n, \nonumber \\
  <0;T|\hat{\phi}^4_0|0;-T>
&=&{1\over \sqrt{T}}\;\sum_{n=0}^{\infty}\;
({-\lambda/ 4})^{n} T^{3 n + 2} \;\; {\bf Q}_n,\nonumber \\
 <0;T| \hat{(\partial_t \phi)}^2_0|0;-T>&
=&{1\over \sqrt{T}}\;\sum_{n=0}^{\infty}\;
 ({-\lambda/ 4})^{n} T^{3 n - 1}  \;\;{\bf P}_n, \nonumber
\end{eqnarray}
with
\begin{eqnarray}
{\bf Z}_n&=&\int_{1>t_n>\ldots t_1>-1} dt_1\ldots dt_n
<0;1|\hat{\phi^4}_{t_n}\ldots
\hat{\phi}^4_{t_1}|0;-1>_0\;,\nonumber\\
{\bf Q}_n&=&\sum_{p=0}^n
\int_{-1<t_0<\ldots <t_{p-1}<0} \int_{0<t_{p+1}
<\ldots <t_n<1}dt_0\ldots \check{dt_p}\ldots
dt_n \nonumber\\
  \;&\;& \times <0;1|\hat{\phi}^4_{t_n}\ldots
\hat{\phi}^4_{t_{p+1
}} \; \hat{\phi}^4_0\;\;\hat{\phi}^4_{t_{p-1}}\,.
\,.\,.\, \hat{\phi}^4_{t_0}|0;-1>_0\;, \nonumber\\
{\bf P}_n&=&\sum_{p=0}^n
\int_{-1<t_0<\ldots <t_{p-1}<0} \int_{0<t_{p+1}<\ldots
<t_n<1}dt_0\ldots \check{dt_p}\ldots dt_n
\nonumber\\
 \;&\;&\times <0;1|\hat{\phi}^4_{t_n}\ldots
\hat{\phi}^4_{t_{p+1}} \;
 \hat{(\partial_t \phi)}^2_0\;
\;\hat{\phi}^4_{t_{p-1}}\ldots
 \hat{\phi}^4_{t_0}|0;-1>_0 \;.\nonumber
\end{eqnarray}
The zero subscripts appended to ket vectors
 mean that the corresponding expectation values
are computed using free propagators with the boundary
conditions discussed in section 2.

In the above expressions, we have extracted
 the $T$ dependences of the coefficients ${\bf Z}_n$,
 ${\bf Q}_n$ and ${\bf P}_n$. In appendix, a
systematic calculation of the ${\bf Z}_n$'s
and ${\bf Q}_n$'s is performed with the help
 of a recursive formula. The values of the
${\bf P}_n$'s are deduced from these
using (14), which gives:
$$ {\bf P}_n=(3n-1/2){\bf Z}_n-2{\bf Q}_{n-1}. $$

 After performing the division by $
Z(T)$  order by order in perturbation,
 the expansion of the mean value of the Hamiltonian
 writes:
\begin{equation}
<\hat{H}_0>^{(p)}_T=\
\sum_{n=0}^p [-{1\over 2}({-\lambda\over 4})^n\,\,
 {\bf P}_n^c\,\, T^{3 n-1}\,\,+\,\,{{\lambda}\over 4}
({-\lambda\over 4})^n\,\, {\bf Q}_n^c\,\, T^{3 n+2}],
\end{equation}
where the $c$ superscripts of ${\bf P}$ and ${\bf Q}$
refer to the contribution of the connected diagrams only.
We have computed these coefficients up
to order 16, and we give here the first few of them:
\begin{equation}\begin{array}{rclrcl}
{\bf Q}_0^c &=&{ 3\over{4}}
                  &{\bf P}_0^c &
=&{ -1\over{2}}\nonumber\\{\bf Q}_1^c &=&{ 33\over{10}}
           &{\bf P}_1^c &=&{ 9\over{10}}\nonumber\\
{\bf Q}_2^c &=&{ 15661\over{700}}
               &{\bf P}_2^c &=&{ 349\over{175}}\nonumber\\
{\bf Q}_3^c &=&{ 3798833\over{19250}}
    &{\bf P}_3^c &=&{ 2363729\over{250250}}\nonumber\\{\bf Q}_4^c &
=&{ 634428694707\over{297797500}}
             &{\bf P}_4^c &=&{ 14511295339\over{223348125}}
\nonumber\\{\bf Q}_5^c &=
&{ 10113181264708\over{372246875}}
            &{\bf P}_5^c &=&{ 8165883862419\over{14145381250}}.
\nonumber\end{array}\end{equation}

\section{The results and their
interpretation.}
Starting from expansion (15) at order $p$, we
perform the replacement $T^{\nu}
\rightarrow \tau^{\nu} (1-\varepsilon)^{-\nu /2}$.
Expanding in powers of $\varepsilon$ up to order
$p-n$, we get the required expansion, which then has to be
optimized with respect to our variational parameter $\tau$:
 \begin{eqnarray}
E^{(p)}_0(\tau)=\sum_{n=0}^p
 [-{1\over 2}({-\lambda\over 4})^n\,\, {\bf P}_n^c
\,\,{\Gamma(p+n/2+1/2)\over \Gamma({3 n}/2+1/2)\,
\Gamma(p-n+1)}
\,\, \tau^{3 n-1}\,\, \nonumber\\ +{{\lambda}
\over 4}({-\lambda\over 4})^n\,\, {\bf Q}_n^c
\,\,{\Gamma(p+n/2+2)\over \Gamma({3 n}/2+2)\,
\Gamma(p-n+1)}\,\, \tau^{3 n+2}].
\end{eqnarray}
 The values of this expression at its optima in $\tau$
 will serve as variational estimates of the value
found in \cite{exact}:
\begin{equation}E^{exact}_0
=\lambda^{1/3} 0.420805\ldots
\end{equation}
Such a polynomial (17) has a number of optima which
increases with the order of perturbation $p$. Most of them
are complex and give a small imaginary part to the
 estimated value of the energy as in the harmonic
approach \cite{bgn}. All of these estimated values
 exhibit the expected $\lambda^{1/3}$ factor.
 We take $\lambda=1$ and compute all the optima up
 to order 16. Results and comments follow in the
next section.

\subsection{Solutions of our variational problem.}

\begin{figure}\vspace{-2cm}
\centerline{\epsffile{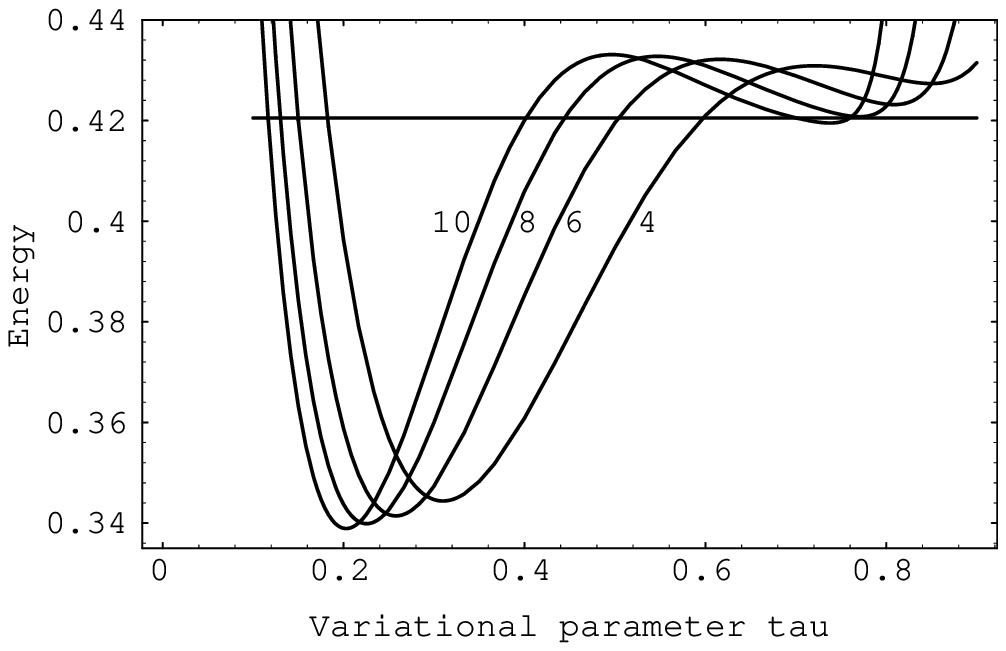}}
\vspace{-1.5cm}\caption{$E_0^{(p)}(\tau)$
 versus $\tau$ for $p=$4, 6, 8 and 10.}
\label{pair}
\end{figure}

\begin{figure}
\centerline{\epsffile{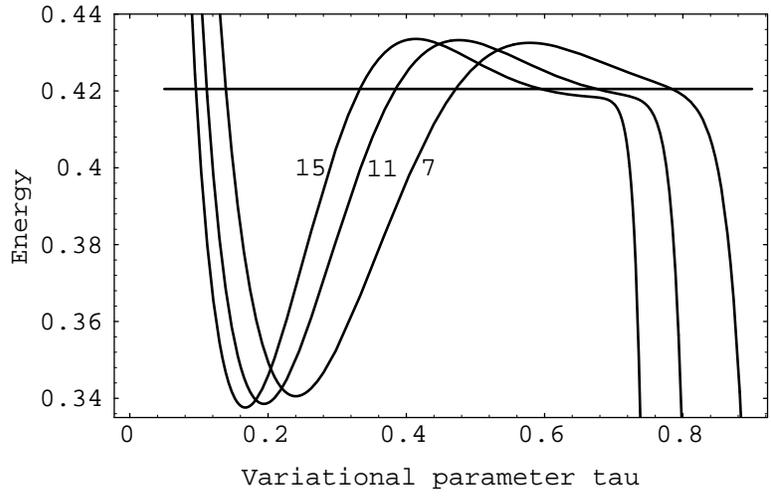}}
\vspace{-1.5cm}
\caption{$E_0^{(p)}(\tau)$ versus $\tau$
for $p=$7, 11 and 15.}
\label{imp}
\end{figure}

$E_0^{(p)}(\tau)$ is plotted in figures \ref{pair} and \ref{imp}
versus the variational parameter for several even and
 odd orders.

Since $\tau$ is
supposed to be irrelevant at infinite  order of
 perturbation, as expected, as $p$ increases,
the curves $E_0^{(p)}(\tau)$ flatten around the
exact value over an increasingly large range
of values of $\tau$ as $p$ increases.

We can follow on these figures the evolution of the
optima with the orders of perturbation.
The first minimum to appear provides a rather poor
estimate ($E_0\simeq 0.34\ldots $)
but  moves to the left of the figure as the order
increases, and is replaced by
 a much more accurate maximum ($E_0\simeq 0.43\ldots $).
It is clear that a third extremum, a minimum, appears
with a still more accurate value $E_0\simeq 0.417\ldots $,
and, from the trend of the odd orders, another, presumably
even better maximum would appear if one would push
the calculation to a couple more orders.

The polynomial  $E_0^{(p)}$ to be optimized has
 an increasing number of stationary points as the
 order of perturbation increases. Plotting real
 optima versus the order of perturbation up to
$p=16$ (figure 3) reveals how they arrange themselves:
 Optimal values of $\tau$ belong to families characterized by the
 straight lines in figure 3 with slopes decreasing
proportionally to the inverse of the square root
of the order $p$. Such a family is created at
first order, followed by another one at  fourth
 order. One reasonably expects other families to
appear at orders greater than $16$, providing
some even more accurate values. Figure 4 plots
the location in $1/\tau^2$ of the
 all the optima up to order
 16  in the first quadrant of the
complex plane. One thus easily follows each
 family up to order 16.

 Such a regular behavior of all these optima is
 explained in the next subsection by considering
the asymptotic expression of $E_0^{(p)}(\tau)$
 for large $p$.

\subsection{The large order behavior.}

For large $p$, using the Stirling formula,
 we find:
\begin{eqnarray}
E^{(p)}_0(\tau)
\;\;\simeq \;\;\sum_{n=0} [-{1\over 2}({-1\over 4})^n
\,\, {\bf P}_n^c\,\,{1\over \Gamma({3 n}/2+1/2)}\,
\, (\tau\,p^{1/2})^{3 n-1}\,\,+ \nonumber\\
{1\over 4}({-1\over 4})^n\,\, {\bf Q}_n^c
\,\,{1\over \Gamma({3 n}/2+2)}\,\, (\tau
\,p^{1/2})^{3 n+2}],
\end{eqnarray}
where we have taken $\lambda=1$. So, at large orders,
the function to be
optimized becomes a series in powers of
 $\tau\,p^{1/2}$. We voluntarily omitted its
upper bound in $n$. Using our knowledge of
 the first coefficients, we establish empirically
the following asymptotic behaviors:
\begin{eqnarray}
{\bf P}_n^c&\sim&{9\over20}
\;\;2^n\Gamma(n),\nonumber\\{\bf Q}_n^c
&\sim&{3\over4}\;\;2^n\Gamma(n+2).\nonumber
\end{eqnarray}
With  these behaviors, the $n^{th}$ term
of the series decreases
like  $1/\sqrt{n!}$ and the series has an infinite
radius of convergence. For large orders $p$, the
search for stationary values
of $E^{(p)}_0(\tau)$ in $\tau$ is nothing but
 the search for those of the series (19) in
 $Y=1/(\tau^2 p)$. Then, real extremal
 values of $Y$ are the slopes of the
families on figure 3. Using the optima
already calculated, we can then extrapolate the
$\tau$ optima and the corresponding
values of the energy for large orders.

 The optima of the real family appearing
 at first order behave asymptotically as:
\begin{equation}1/(\tau^{opt})^2\;\simeq
\;1.1612\;+\; 2.31156\;p\;+\;{0.025\over p}\;.
\end{equation}
 The corresponding value for the energy tends toward
 a rather bad approximation from above:
\begin{equation}
E_0^{opt}\;\simeq \; 0.335
\;+\;{0.038\over p}\;.
\end{equation}

 For the second real family which appears at the
fourth order
\begin{equation}
1/(\tau^{opt})^2
\;\simeq\;0.4651\;+\; 0.3578\;p\;+\;{0.1245\over p}\;,
\end{equation}
the corresponding energy tends towards
 a better value from below:
\begin{equation}
E_0^{opt}\;\simeq\; 0.434\;-\;{0.014\over p}\;.
\end{equation}

The asymptotic value provided by a family is more accurate
 when the family appears later in perturbation. We
also verify that the slopes tend to decrease. This
corresponds to the successive extrema of (22) appearing
at smaller and smaller values of $Y$, corresponding to those
of the function $E^{as}_0(Y)$

\begin{eqnarray}
E^{as}_0(Y)\;=\;\sum_{n=0}^{\infty}
 [-{1\over 2}({-1\over 4})^n\,\, {\bf P}_n^c\,\,{1\over
 \Gamma({3 n}/2+1/2)}\,\, Y^{-(3 n-1)/ 2}\,\,+
 \nonumber\\{1\over 4}({-1\over 4})^n\,\,
{\bf Q}_n^c\,\,{1\over \Gamma({3 n}/2+2)}\,\,
Y^{-(3 n+2)/ 2}]\;.
\end{eqnarray}
Using the available coefficients up to order 16
 is sufficient to see that this function has a minimum:
$$\bar Y\;\simeq\;2.2629
\;\;\;\; ;\;\;\;\;\;E^{as}_0(\bar Y)\;\simeq \;0.3315.$$
One recognizes the slope and the asymptotic
energy of the first real family within a few percents.
Using the empirical asymptotic behaviors of the
coefficients ${\bf P}$ and ${\bf Q}$ would extend the
range of $Y$ where one could compute $E^{as}_0$, revealing
the asymptotic properties of the other families.

\section{Conclusion}
The results presented in this paper are qualitatively
very similar to those obtained in the
harmonic approach \cite{bgn}: In the same way as
in the harmonic approach,
 the optima arrange themselves
 in families which are understood using a large order
behavior. The set of these families provide a
sequence of approximate values which converges to
 the exact one. However, for a given perturbative
 order, our estimated values were more accurate
in the harmonic approach. Indeed, for the first
family, we obtain an estimated value of the
ground state energy with a precision of $2.10^{-5}$
 in the harmonic case and of $2.10^{-1}$ in the
present approach.This is presumably due to several
causes: in the compact time case, there is no time
translation invariance, hence no Hamiltonian,
and we only compute the expectation value of an
operator which becomes the Hamiltonian in the
large time box limit.
Furthermore, as explained in section 2, our
perturbative expansions do not take into
account the complete $\phi^4$ interaction term
 but only reconstruct it more and more
accurately as the perturbative order
 increases. Finally, a compact Euclidean time
is clearly quite far from the physics of an
anharmonic oscillator, much farther than
a harmonic approximation. One may even consider it quite
remarkable that the procedure nevertheless seems
to converge in the same manner, albeit much more slowly.
One could thus contemplate using such a compactification,
which could involve only the spatial coordinates
or both space and time coordinates as a gauge invariant
variational approach to gauge theories.

\section{Appendix: Calculation of
 coefficients.}
\subsection{Calculation of the ${\bf Z}_n$'s}
\begin{eqnarray}  <0;T|0;-T>&=& \sum_{n=0}^{\infty}
 ({-\lambda\over 4})^n T^{3 n} {\bf Z}_n\nonumber\\
{\bf Z}_n&=&\int_{-1<t_1<\ldots <t_n<1} dt_1\ldots dt_n
 <0;1|\hat{\phi}^4_{t_n}\ldots
\hat{\phi}^4_{t_1}|0;-1>_0\nonumber
\end{eqnarray}

For each time $t_1,t_2,\ldots ,t_n$ , one
 uses the closure relation in
position space. The $\hat{\phi}$
 operator is diagonal in this basis.
 One thus rewrites the integrand in
the following way: $$<0;
1|\hat{\phi}^4_{t_n}\ldots \hat{\phi}^4_{t_1}|0;-1>_0
\;=   \int^{\infty}_{-\infty}dx_n\ldots dx_1
\;<0;1\mid x_n;t_n>_0\;x^4_n $$
$$\times <x_n;t_n\mid \ldots
\mid x_1;t_1>_0\;x^4_1\;<x_1;t_1\mid 0;-1>_0.$$

For $t'>t$, the free propagator writes: $$
<y;t'|x;t>={1\over \sqrt{ 2 \pi (t'-t)}}
\exp{-\frac{(y-x)^2}{2(t'-t)}}.\nonumber$$
This product of expectation values of operators
is a Gaussian function in terms of $x_1,x_2,\ldots,
x_n$.One thus performs  integrals over
 all positions simultaneously to obtain:
\begin{eqnarray}
<0;1|\hat{\phi}^4_{t_n}\ldots
\hat{\phi}^4_{t_1}|0;-1>_0\;&=&{1\over 2
\sqrt\pi }\;\;d^4_{j_1}\ldots d^4_{j_n}\;
\;\exp({1\over4}J^T.\,A.\,J)\mid_{J\equiv0}
\nonumber\\&=&{1\over 2 \sqrt\pi }\;\;d^4_{j_1}
\ldots d^4_{j_n}\;\;{1\over(2 n)!}{1\over4^{2 n}}
\;(J^T.\,A.\,J)^{2 n}\;, \nonumber
\end{eqnarray}
where
\begin{eqnarray}
A_{ij}&=&(t_i+1)(1-t_j)
\;\; \mbox{for}\;\; j\geq i \;\;\; (t_j\geq  t_i),
\nonumber \\ {\rm det}\,A&=&2^{n-1}
\;\prod_{i=0}^n\;(t_{i+1}-t_i).\nonumber
\end{eqnarray}
$d^4_{j_n}$ is the fourth derivative
with respect to $j_n$. Now, we build a recursive
 procedure which  performs the four derivatives
with respect to $j_n$ ($d^4_{j_n}$) together with
 the integral over $t_{n}\in [t_{n-1};1]$. To do
so, one will use the following identity:
$$ \int_{t'}^1 dt(1+t)^{\alpha}(1-t)^{\beta}={1
\over1+\alpha+\beta}{1\over C_{\alpha+\beta}^{
\beta}}\;\sum_{j=0}^{\alpha}\;2^j\, C_{\alpha+
\beta-j}^{\beta}\;(1+t')^{\alpha-j}(1-t')^{\beta+1},$$
with
$$C_n^p=\frac{n!}{p!(n-p)!}.$$
Let us define:
$$\Omega(n,m,\alpha,\beta)\,=\,\int_{-1}^1dt_1\ldots
\int_{t_{N-1}}^1dt_N\;d^4_{j_1}\ldots d^4_{j_N}\;
( \underbrace{J.A.J}_{N})^n\;( \underbrace{J.A}_{N})^m
\mid_{J\equiv0}$$
$$\times (1+t_N)^{\alpha}(1-t_N)^{\beta}\mid_{J\equiv0}$$
\begin{eqnarray}
 =4!\int_{-1}^1dt_1
\ldots \int_{t_{N-2}}^1&&\mbox{\hspace{-6mm}}
dt_{N-1}\;d^4_{j_1}\ldots d^4_{j_{N-1}}\;C_n^i
\;(\sum_{k=0}^i\;C_m^{4-i-k}\,C_i^k\,2^{i-k})\;
(\underbrace{J.A.J}_{N-1})^{n-i}\nonumber\\
\;&\,&\mbox{\hspace{-6mm}}
\;\times (\underbrace{J.A}_{N-1})^{m+2i-4}\;
(1+t_{N-1})^{4-i+\alpha}(1-t_{N-1})^{i+\beta}
\mid_{J\equiv0}\nonumber
\end{eqnarray}
where:$$\underbrace{J.A.J}_{N}\;=\;
\sum_{i,k=1}^N\,j_iA_{ik}j_k\;\;\;
\mbox{and}\;\;\underbrace{J.A}_{N}
\;=\;\sum_{i=1}^N\,j_i\,(t_i+1).$$
So that $\Omega$ satisfies:
\begin{eqnarray}
\mbox{\hspace{-6mm}} \Omega(n,m,&\alpha&,\beta)
\;=\;4!\sum_{i=0}^4\Lambda_i(n,m)
\sum_{j=0}^{4+\alpha-i}\Theta_{i,j}(\alpha,\beta)
\,\;\;\; \nonumber \\
&\times&\Omega(n-i,m+2i-4,4+\alpha-i-j,i+\beta+1)
\end{eqnarray}
$$ \Lambda_i(n,m)\,=\,C_n^i\;
\sum_{k=0}^i\;C_m^{4-i-k}\,C_i^k\,2^{i
-k}\;$$
$$\Theta_{i,j}(\alpha,\beta)\,=
\,{C_{4+\alpha+\beta-j}^{i+\beta}
\over C_{4+\alpha+\beta}^{i+\beta}}\;
{2^j\over5+\alpha+\beta}$$
with the boundary conditions:
$$ \Omega(n,m,\alpha,\beta)=0\;\;\;\;\mbox{as soon
 as $n$ or $m$ is negative,} $$
$$ \Omega(0,0,\alpha,\beta)=\lim_{t\rightarrow -1}(1+t)^{
\alpha}(1-t)^{\beta}=\delta(\alpha)\;2^{\beta}, $$
for the last integration to be performed
between $-1$ and $1$. The coefficients ${\bf Z}_n$
 of the perturbative expansion are:
$${\bf Z}_n\;={1\over 2 \sqrt{\pi}}{1
\over (2n)!}{1\over 4^{2n}}\;\Omega(2 n,0,0,0).$$

\subsection{Calculation of the ${\bf Q}_n$'s.}
\begin{eqnarray}  <0;T|\hat{\phi}^4_0|0;-T>&=
&{1\over \sqrt{T}} \; \sum_{n=0}^{\infty} ({-
\lambda\over 4})^{n} T^{3 n+2} {\bf Q}_n\nonumber\\
{\bf Q}_{n-1}&=&\sum_{p=1}^n\;{\bf Q}_{n-1}^p\nonumber\\
{\bf Q}_{n-1}^p&=&\int_{-1<t_1<\ldots <t_{p-1}
<0<t_{p+1}<\ldots <t_n<1} dt_1\ldots \check{dt_p}\ldots
dt_n\nonumber\\&\,&\times <0;1|\hat{\phi}^4_{t_n}
\ldots \hat{\phi}^4_{t_{p+1}}\hat{\phi}^4_{0}
\hat{\phi}^4_{t_{p-1
}}\ldots \hat{\phi}^4_{t_1}|0;-1>_0\nonumber
\end{eqnarray}
The preceding method must now
be adapted to a slightly different domain of
integration over the times (now $t_p=0$). We
will consider four different regions in order
 to build the recursive definition of a new
function $\Omega_p(n,m,\alpha,\beta)$;
$$\underbrace{\int_{-1}^0dt_1\;.\;.\;.\;
\int_{t_{p-2}}^0dt_{p-1}}_{(4)}\;\;
\underbrace{``(t_p=0)"}_{(3)}\;\;
\underbrace{\int_{0}^1dt_{p+1}}_{(2)}\;\;
\underbrace{\;.\;.\;.\;
\int_{t{n-1}}^1dt_n}_{(1)}
\;\;\;\;\;\;\;\;\;\;$$
These four different domains of integration
 can enter the preceding frame.The quantity
 $(2 n+m)/4=n,\ldots ,1$ counts the remaining
iterations to be performed and decreases by one
 unit at each step. One then builds the
 $\Omega_p$'s  for each of these
steps, writing:
\begin{eqnarray}&n'=n-i
\;\;\;,\;\;\;&m'=m+2i-4,\nonumber\\
&\alpha'=\alpha-i-j\;\;\;,\;\;\;
&\beta'=i+\beta+1.\nonumber
\end{eqnarray}
Sums over $i$ and $j$ are respectively
performed between $0$ and $4$, and $0$
and $4+\alpha-i$.
\paragraph{1. When $(2n+m)/4$
 is greater than $p+1$:} $\Omega_p$ is
 just as $\Omega$:
$$\Omega_p(n,m,
\alpha,\beta)\;=\;\sum_i\;\sum_j\;\Lambda_i(n,m)
\;\Theta_{i,j}(\alpha,\beta)\;\Omega_p(n',m',
\alpha',\beta').$$
\paragraph{2. When $(2n+m)/4$
equals $p+1$:}Here $t_p$, the lower bound of the
integral is driven to zero:
$$(1+t_p)^{\alpha'}
\;(1-t_p)^{\beta'}\mid_{t_p=0}\;=\;1\;=\;(1+t_p)^0
\;(1-t_p)^0$$
Then, choosing $\alpha',\beta'=0$,
 one can incorporate this case in the preceding frame:
$$\Omega_p(n,m,\alpha,\beta)\;=\;\sum_i\;\sum_j
\;\Lambda_i(n,m)\;\Theta_{i,j}(\alpha,\beta)\;
\Omega_p(n',m',0,0)$$
\paragraph{3. When $(2n+m)/4$
 equals $p$:}At this step, no integral has to be
 taken over $t_p$ so that there is no sum
 over $j$ and the exponents $\alpha'$ and
 $\beta'$ remain zero:
$$\Omega_p(n,m,0,0)
\;=\;\sum_i\;\;\Lambda_i(n,m)\;\Omega_p(n',m',0,0)$$
\paragraph{4. When $(2n+m)/4$ is smaller than $p$:}
The domain of integration is obtained subtracting
 case (2) from  case (1):
$$\int_{t_{i-1}}^0dt_i\;=\;\int_{t_{i-1}}^1dt_i\;-\;
\int_{0}^1dt_i\;\;;$$
\begin{eqnarray}\Omega_p(n,m,\alpha,\beta)
\;&=&\;\sum_i\;\sum_j\;\Lambda_i(n,m)\;\Theta_{i,j}(
\alpha,\beta)\;\nonumber\\&\,&\times
(\Omega_p(n',m',\alpha',\beta')\;-
\;\Omega_p(n',m',0,0)).\nonumber
\end{eqnarray}
This together with the boundary conditions
$$\Omega_p(n,m,\alpha,\beta)=0\;\;\;\;\mbox{as
 soon as $n$ or $m$ is negative,} $$
$$ \Omega_p(0,0,\alpha,\beta)=\delta(
\alpha)\;2^{\beta}. $$
 defines $\Omega_p$ for every $p$.

The coefficients ${\bf Q}_{n-1}$ are then:
$${\bf Q}_{n-1}\;={1\over 2
\sqrt{\pi}}{1\over (2n)!}{1\over 4^{2n}}\;\sum_{p=1}^n
\;\Omega_p(2 n,0,0,0).$$
The values of the ${\bf P}_n$'s
are then deduced using formula (16) which writes in
terms of the other coefficients:
\begin{equation}
{\bf P}_n\;=\;(3\,n\,-\,1/2)\;{\bf Z}_n\;-\:2\;{\bf Q}_{n-1}\;
\;\;\forall\,n.
\end{equation}

\section{Acknowlegments}

One of us (A. N.) is grateful for their
hospitality to the  Lawrence Berkeley Laboratory,
where this work was supported in part by the Director,
Office of Energy research, Office of High Energy and
Nuclear Physics, Division of High Energy Physics of
the U.S. Department of Energy under Contract
DE-AC03-76SF00098 and to the Physics Department,
University of California,
Berkeley, where this work was completed with partial
support from National Science Foundation
grant PHY-90-21139.

\newpage

\begin{figure}
\vspace{-2cm}\centerline{\epsffile{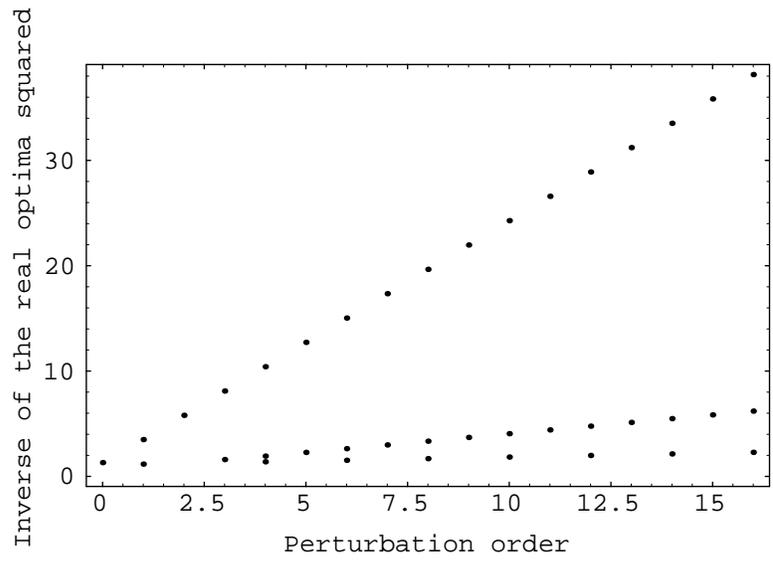}}
\vspace{-1cm}\caption{Real optima  $(1/\tau^{opt})^2$
 versus the order $p$. }
\label{retau}
\end{figure}

\begin{figure}
\centerline{\epsffile{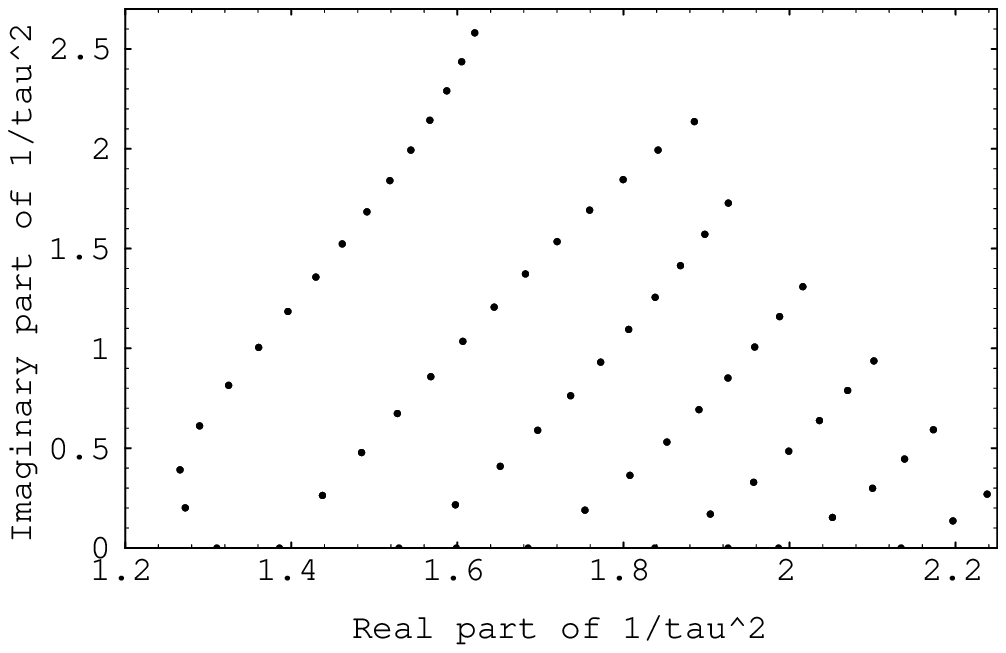}}
\vspace{-1.5cm}\caption{All the optima
 $(1/\tau^{opt})^2$ in the complex plane
 up to order 16. }
\label{otau2}
\end{figure}


\begin{thebibliography}{aaaaaaa}

\bibitem{nev}
A. Neveu, Nucl. Phys. B 18 (1990) 242.

\bibitem{bgn}B. Bellet, P. Garcia and A. Neveu,
``Convergent sequences of perturbative approximations for the
anharmonic oscillator, I. Harmonic approach",
Montpellier preprint, preceding paper.

\bibitem{agn}C. Arvanitis, F. Geniet and A. Neveu,
``Variational solution of the Gross-Neveu model, I. The large-$N$ limit",
Montpellier preprint PM94-19 (hep-th/9506188).

\bibitem{MIT}
A. Chodos, R.L. Jaffe, K. Johnson, C.B. Thorn and V.F. Weisskopf,
Phys. Rev. D 9 (1974) 3471;\\
A. Chodos, R.L. Jaffe, K. Johnson and C.B. Thorn,
Phys. Rev.  D 10 (1974) 2599;\\
K. Johnson, Lectures presented at the Scottish Universities
Summer School, August 1976, St. Andrews, Scotland;\\
For a review: P. Hasenfratz and J. Kuti, Phys. Rep. 40
(1978) 75.

\bibitem{LF}
M. L\"uscher, Phys. Lett. 118 B (1982) 391;\\
M. L\"uscher, DESY preprint 83-116 (1983);\\
M. L\"uscher, Nucl. Phys. B 219 (1983) 233;\\
E.G. Floratos, Phys. Lett. 154 B (1985) 173;\\
E.G. Floratos and D. Petcher, Phys. Lett. 160 B (1985) 271;\\
E.G. Floratos and N.D. Vlachos, Phys. Lett. 189 B (1987) 137;\\
E.G. Floratos, N.D. Tracas and N.D. Vlachos,
Phys. Lett. 253 B (1991) 399.

\bibitem{exact}C.M. Bender,
 K. Olaussen and P.S. Wang, Phys. Rev.  D 16 (1977) 1780.

\end{thebibliography}
\end{document}